\documentclass[aps,prl,twocolumn,showpacs,nofootinbib]{revtex4}

\begin{document}

%%% article in Russian
%\rus

%%% article title
\title{Upper limit for superconducting transition temperature
in Eliashberg -- McMillan theory}

%%% article title - for colontitle (at the top of the page)
%\rtitle{Hall effect in doped Mott insulator}

%%% article title - for table of contents (usually identical with \title)
%\sodtitle{Hall effect in doped Mott insulator}

%%% author(s) ( + e-mail)

\author{M.\ V.\ Sadovskii}

%%% author's address(es)
\affiliation{Institute for Electrophysics, RAS Ural Branch, Amundsen st. 106,
Ekaterinburg 620016, Russia}

\begin{abstract}
We present simple qualitative estimates for the maximal
superconducting transition temperature, which may be achieved due to
electron -- phonon coupling in Eliashberg -- McMillan theory.
It is shown that in the limit of very strong coupling the upper limit for
transition temperature is determined in fact by a combination of atomic
constants and density of conduction electrons.
\end{abstract}

\pacs{74.20.-z}

\maketitle

Experimental discovery of high -- temperature superconductivity in hydrides
under high (megabar) pressures \cite{H3S,ErPR} stimulated the search for the ways
to achieve superconductivity at room temperature \cite{Pud}. At the moment the
common view \cite{Grk-Krs,WPick} is  that the high -- temperature
superconductivity in hydrides can be described in the framework of the standard
Eliashberg -- McMillan theory \cite{Elias-1,Elias-2,McM}. Within this theory
many attempts were undertaken to estimate the maximal achievable 
superconducting transition temperature and the discussion of some of these 
attempts can be found in the reviews \cite{Grk-Krs,WPick,MS}. 
In the recent paper \cite{Trach} a new upper limit for $T_c$ was proposed, 
expressed as some combination if fundamental constants. Below we shall show
that with minor modifications such $T_c$ limit follows directly from
Eliashberg -- McMillan theory.

Traditionally, after the appearance of BCS theory, in most papers devoted to
possible ways of increasing $T_c$, discussion develops in terms of dimensionless
constant of electron -- phonon coupling $\lambda$ and characteristic (average)
frequency $\langle\Omega\rangle$ of phonons, responsible for Cooper
pairing. In their fundamental paper \cite{AD} Allen and Dynes obtained in the 
limit of very strong coupling $\lambda>10$ the following expression for 
$T_c$\footnote{In fact this asymptotic behavior works rather satisfactorily
already for $\lambda>2$}:
\begin{equation}
T_c = 0.18\sqrt{{\lambda}\langle\Omega^2\rangle}
\label{AD_asymp}
\end{equation}
Then it seems that limitations for the value of $T_c$ are just absent, so that 
quite high values of $T_c$ can be obtained with electron -- phonon pairing
mechanism. In reality the situation is more complicated. Actually 
parameters $\lambda$ and $\langle\Omega^2\rangle$ in Eliashberg -- McMillan
theory are not independent, which is well known for quite a time
\cite{Grk-Krs,WPick,MS,Max}.

The relation of $\lambda$ and $\langle\Omega^2\rangle$ is clearly expressed by
McMillan's formula for $\lambda$, first derived in Ref. \cite{McM}:
\begin{equation}
\lambda=\frac{N(0)\langle I^2\rangle}{M\langle\Omega^2\rangle}
\label{McM_form}
\end{equation}
where  $M$ is an ion mass, $N(0)$ is electronic density of states at the
Fermi level and we introduced the matrix element of the gradient of
electron -- ion potential, averaged over the Fermi surface:
\begin{eqnarray}
\langle I^2\rangle=
%\frac{1}{[N(0)]^2}\sum_{\bf p}\sum_{\bf p'}
%\left|I({\bf p-p'})\right|^2\delta(\varepsilon_{\bf p})
%\delta(\varepsilon_{\bf p'})=\nonumber\\
\frac{1}{[N(0)]^2}\sum_{\bf p}\sum_{\bf p'}
\left|\langle{\bf p}|
\nabla V_{ei}({\bf r})|{\bf p'}\rangle\right|^2)\delta(\varepsilon_{\bf p})
\delta(\varepsilon_{\bf p'})=\nonumber\\
=\langle |\langle {\bf p}|\nabla V_{ei}({\bf r})|{\bf p'}\rangle|^2\rangle_{FS}\nonumber\\
\label{grV2}
\end{eqnarray}
Here $\varepsilon_{\bf p}$ is the spectrum of free electrons, with energy zero
chosen at the Fermi surface. Eq. (\ref{McM_form}) gives very
useful representation for the coupling constant $\lambda$, which is
routinely used in the literature and in practical (first -- principles)
calculations \cite{WPick}.

Using Eq. (\ref{McM_form}) in Eq. (\ref{AD_asymp}) we immediately obtain:
\begin{equation}
T_c^{\star}=0.18\sqrt{\frac{N(0)\langle I^2\rangle}{M}}
\label{AD_McM}
\end{equation}
so that both $\lambda$ and $\langle\Omega^2\rangle$ just drop out from the expression
for $T_c^{\star}$, which is now expressed via Fermi surface averaged matrix element
of electron -- ion potential, ion mass and electron density of states at the
Fermi level. The only deficiency of this expression is the loss of intuitive
understanding due to the absence of parameters in terms of which $T_c$ is
usually treated.

As was already noted, all parameters entering this expression can be rather 
simply obtained during the first -- principles calculations of $T_c$  for
specific materials (compounds) \cite{WPick}. Let us also stress that the value
of $T_c^{\star}$ defined in Eq. (\ref{AD_McM}), calculated for any specific
material does not have any direct relation to real value of $T_c$, but just
defines precisely the upper limit of $T_c$, which ``would be achieved'' in the 
limit of strong enough electron -- phonon coupling. Below we shall present some
elementary qualitative estimates of its value.

In the following we shall assume to be dealing with three -- dimensional metal
with cubic symmetry with an elementary cell with lattice constant
$a$ and just one conduction electron per atom. Then we have:
\begin{equation}
N(0)=\frac{mp_F}{2\pi^2\hbar^3}a^3
\label{DOS}
\end{equation}
where $p_ F\sim\hbar/a$ is the Fermi momentum, $m$ is the mass of free (band) 
electron. Electron -- ion potential (single -- charged ion, $e$ is electron 
charge) can be estimated as:
\begin{equation}
V_{ei}\sim\frac{e^2}{a}\sim e^2p_F/\hbar
\label{Vei}
\end{equation}
so that its gradient is:
\begin{equation}
\nabla V_{ei}\sim\frac{e^2}{a^2}\sim e^2p_F^2/\hbar^2
\label{gradV}
\end{equation}
Then we easily obtain the estimate of (\ref{grV2}):
\begin{equation}
I^2\sim \left(\frac{e^2}{a^2}\right)^2\sim (e^2p_F^2/\hbar^2)^2
\label{mxel}
\end{equation}
Here we have dropped different numerical factors of the order of unity.
Collecting them back in the model of free electrons we get an estimate for
$T_c^{\star}$ from Eq.  (\ref{AD_McM}) as:
\begin{equation}
T_c^{\star}\sim 0.2\sqrt{\frac{m}{M}}\frac{e^2}{\hbar v_F}E_F
\label{Tstar}
\end{equation}
where $E_F=p_F^2/2m$ is Fermi energy, $v_F=p_F/m$ is electron velocity at the 
Fermi surface. The value of $\frac{e^2}{\hbar v_F}$, as is well known, 
represents the dimensionless coupling for Couloumb interaction and for
typical metals it is of the order of or greater than unity. The factor of
$\sqrt{\frac{m}{M}}$ determines isotopic effect.

Let us measure length in units of Bohr radius $a_B$ introducing the standard
dimensionless parameter $r_s$ by relation $a^3=\frac{4\pi}{3}(r_sa_B)^3$.
Then we have:
\begin{equation}
a\sim r_sa_B=r_s\frac{\hbar^2}{me^2}=r_s\frac{\hbar}{mc\alpha}
\label{a_b}
\end{equation}
where we have introduced the fine structure constant
$\alpha=\frac{e^2}{\hbar c}$. Correspondingly the Fermi momentum is given by:
\begin{equation}
p_F\sim\frac{\hbar}{r_sa_B}=\frac{me^2}{\hbar r_s}=\frac{mc}{\hbar r_s}\alpha
\label{PFer}
\end{equation}
Then $T_c^{\star}$ (\ref{AD_McM}) can be rewritten as:
\begin{equation}
T_c^{\star}\sim\frac{0.2}{r_s}\sqrt{\frac{m}{M}}\alpha^2mc^2/2\sim
\frac{0.2}{r_s}\sqrt\frac{m}{M}\frac{me^4}{2\hbar^2}\sim
\frac{0.2}{r_s}\sqrt\frac{m}{M}Ry
\label{Tc_star}
\end{equation}
where $Ry=me^4/2\hbar^2\approx$ 13.6 eV is the Rydberg constant.
Here we have obtained the same combination of fundamental (atomic) constants,
which was suggested in Ref. \cite{Trach}, by some quite different reasoning,
as determining the upper limit of superconducting critical temperature.
However, our expression contains an extra factor of $r_s^{-1}$, which
necessarily reflects the specifics of a material under consideration
(density of conduction electrons), so that the value of $T_c^{\star}$ is in no
sense universal.

As was already noted above the value of $T_c^{\star}$ strictly speaking has no
relation at all to the real superconducting transition temperature  $T_c$.
However, expressions (\ref{Tstar}) and (\ref{Tc_star}) may be useful to
estimate ``potential perspectives'' of some material in the sense of achieving
high values of transition temperatures under the conditions of strong
electron -- phonon coupling.
For example in metallic hydrogen $M$ is equal to proton mass and we have
$\sqrt\frac{m}{m_p}\sim 0.02$, so that for $r_s=1$ we get an estimate of
$T_c^{\star}\sim 650$ K. This is in nice agreement with the result of
$T_c=$ 600 K, obtained in Ref. \cite{Max} solving Eliashberg equations
for FCC lattice of metallic hydrogen with $r_s=1$, taking into account the
calculated softening of the phonon spectrum, leading to realizations of very
strong coupling ($\lambda=6.1$). 
At the same time in the recent paper \cite{vdM} an elegant numerical study
of superconductivity of metallic hydrogen within ``jellium'' model has shown,
that the maximal value of $T_c$ can be achieved at $r_s\sim 3$, not
exceeding 30 K. This is obviously related to the fact that in the ``jellium''
model the weak coupling is realized and there is no softening of the phonon
spectrum. Finally we hope that Eqs. (\ref{Tstar}) and (\ref{Tc_star}) can be
relevant for preliminary estimates of $T_c$ in some of the metallic hydrides,
which are currently under intensive study in the search for room -- temperature
superconductivity.

%\newpage

\end{document}